\documentclass[12pt]{article}

\usepackage{graphicx}
\usepackage{dcolumn}
\usepackage{bm}
\usepackage{amsmath}
\usepackage{epstopdf}
\usepackage{amsfonts}
\usepackage{amssymb}
\usepackage{epsfig}
\usepackage{tabularx}
\usepackage{color}
\usepackage{hyperref}

\begin{document}

\begin{titlepage}



\centerline{\large \bf {Greybody factors for a nonminimally coupled scalar field}}

\vskip 0.2 cm

\centerline{\large \bf {in BTZ black hole background}}

\vskip 1cm

\centerline{G. Panotopoulos$^1$ and {\'A}ngel Rinc{\'o}n$^2$}

\vskip 1cm

\centerline{CENTRA, Instituto Superior T{\'e}cnico, Universidade de Lisboa}

\vskip 0.2 cm

\centerline{Av. Rovisco Pais 1, Lisboa, Portugal$^1$}

\vskip 0.2 cm

\centerline{Instituto de F{\'i}sica, Pontificia Universidad Cat{\'o}lica de Chile}

\vskip 0.2 cm

\centerline{Av. Vicu{\~n}a Mackenna 4860, Santiago, Chile$^2$}

\vskip 0.5 cm

\centerline{email:
\href{mailto:grigorios.panotopoulos@tecnico.ulisboa.pt}{\nolinkurl{grigorios.panotopoulos@tecnico.ulisboa.pt}}, \href{mailto:arrincon@uc.cl}{\nolinkurl{arrincon@uc.cl}}}

\begin{abstract}
In the present work we study the propagation of a probe nonminimally coupled scalar field in BTZ black hole background. We find analytical expressions for the reflection coefficient, the absorption cross-section, and the decay rate in the strong coupling case, where the nonminimal coupling is larger than its conformal value  $\xi_{c}= 1/6$. Our main results are summarized in several figures, in which we show how the behavior of the aforementioned quantities depends on the parameters of the theory. Our findings show that i) the reflection coefficient tends to zero only for a nonminimal coupling $\xi \geq 0.25$, and ii) in the zero angular-momentum case the greybody factor in the low energy regime tends to a finite constant that generically does not coincide with the area of the horizon of the black hole. There is, however, a special case in which this holds.
\end{abstract}


\end{titlepage}

\section{Introduction}

After Hawking's seminal papers where he showed that black holes emit radiation from the horizon \cite{hawking1,Hawking:1974sw}, black-holes
have become exciting objects and an excellent laboratory to study and understand quantum gravity. It is often said that
the Hawking radiation is a black body radiation, thermal in nature, characterized entirely by the hawking temperature $T_H$. This,
however, is only partially true. The reason why is that the emitted particles feel an effective potential barrier in the exterior
region. The potential barrier backscatters a part of the outgoing radiation back into the black hole \cite{kanti1}. The greybody factor, or
else absorption cross section $\sigma_{abs}(\omega)$, is a frequency dependent factor that measures the modification of the original
black body radiation. To see that we make use of the black hole differential decay rate into bosons of fermions of energy $\omega$ given by
the Hawking formula \cite{Hawking:1974sw,kanti1}
\begin{equation}\label{energyrate}
\frac{dE(\omega)}{dt} = \sum_{\ell} \sigma_{\ell}(\omega)\frac{\omega}{e^{\omega/T_H} \mp 1} \frac{d^3 k}{(2 \pi)^3}
\end{equation}
The total black hole emission rate is obtained by integrating the greybody factor $\sigma_{\ell}$ over
all spectra. Moreover, if the greybody factor is a constant the black hole emission
spectrum would be exactly that of a blackbody radiation. This is the non-triviality of the greybody factor which leads
to deviations of blackbody emissions and the consequent greybody radiation \cite{Gonzalez:2010ht}.

Greybody factors are important both from the theoretical and from the experimental point of view \cite{kanti1}. From the theory side, they
give us valuable information about the near horizon structure of black holes. From the experiment side, although the Hawking radiation has
not been detected yet, is that greybody factors modify the spectrum in the region where most particles are produced. This could be essential
in studying the collider signatures of the evaporation of TeV mini black-holes \cite{mini}.

Relativistic scattering of waves has been traditionally studied in asymptotically flat spacetimes without a cosmological constant \cite{Doran:2005vm,dolan2006fermion,crispino2007absorption,Dolan:2008kf,crispino2009electromagnetic,crispino2009scattering,Crispino:2009zza}.
However, due to inflation \cite{guth}, the current cosmic acceleration \cite{riess,Perlmutter:1998np} and the AdS/CFT
correspondence \cite{adscft,Klebanov:2000me}, asymptotically non-flat spacetimes with a positive or negative cosmological constant
have also been studied over the years \cite{3D,Myung:2002is,Fernando:2004ay}. Of particular interest is the Ba{\~n}ados, Teitelboim and
Zanelli black-hole (BTZ) \cite{BTZ,Banados:1992gq}, which lives in three dimensions, and the presence of a negative cosmological constant
is crucial for the existence of the black-hole. A complete review on BTZ black-hole can be found in \cite{review}. Gravity in (1+2) dimensions
is special for several reasons. First, due to the absence of propagating degrees of freedom the treatment is simpler. Second, because it
has thermodynamic properties closely analogous to those of realistic (1+3)-dimensional black holes: it radiates at a Hawking temperature \cite{review}.
Additionally, the Einstein-Hilbert action is closely related to a Yang-Mills theory with only the
Chern-Simons term \cite{CS,Witten:1988hc,Witten:2007kt}.

Regarding greybody factors, another interesting aspect is Universality proved in \cite{universality}, which states that in a generic
spherically symmetric spacetime and in any number of dimensions, when the angular-momentum vanishes the greybody factor for a
minimally coupled massless scalar field in the low energy regime goes to the area of horizon. The greybody factors for a massive scalar
field were analyzed in \cite{massive}, and it was shown there that Universality is respected only under certain restrictions of the parameters
involved. Lately, there is an interest in studying and analysing the greybody factors for scalar fields with a nonzero coupling to the scalar
curvature \cite{coupling,kanti2,Pappas:2016ovo,Ahmed:2016lou}. In particular, in \cite{coupling} it was shown that the greybody factor in the
low energy regime and in the zero angular-momentum case tends to zero like $\omega^2$.

Despite the previous works, to the best of our knowledge the greybody factors for a nonminimally coupled scalar field in BTZ spacetime is
still lacking. Given the importance of (1+2) gravity in general, and of the BTZ black-hole in particular, combined with the increasing
interest in nonzero coupling to the scalar curvature, we wish in this work to find analytical expressions for the reflection coefficient,
the absorption cross-section and the decay rate for a nonminimally coupled scalar field in BTZ background. Our work
is organized as follows: After this introduction, we present the classical BTZ black hole solution in the next section,
while the nonminimally coupled scalar field and its wave equation are discussed in section 3, in which we also show the
effective potential barrier. In the fourth section we solve the radial equation analytically in terms of hypergeometric functions, and we
compute the reflection coefficient as well as the absorption cross section and the
decay rate in section 5. Finally, we conclude our work in the last section.

\section{Classical gravity in (1+2) dimension}\label{clasico}

Our starting point is the Einstein-Hilbert action with cosmological constant, which, in (1+2) dimension reads
\begin{align}
S_0 &= {1 \over 2\kappa } \int {\mathrm {d}}^{3}x {\sqrt {-g}}\,
\Bigl[R-2\Lambda\Bigl]
\end{align}
where $\kappa = 8 \pi G$ is the Einstein's constant, $R$ is the Ricci scalar and $\Lambda=-1/l^2$ is the cosmological constant.
The classical equation of motion is given by the Einstein field equation
\begin{align}
G_{\mu \nu} + \Lambda g_{\mu \nu } &= \kappa T_{\mu \nu }
\end{align}
We now consider the well known 3-dimensional BTZ black hole \cite{BTZ,Banados:1992gq} without angular momentum ($J=0$, diagonal metric) and mass $M$, which line element is
\begin{equation}
ds^2 = -f(r) dt^2 + f(r)^{-1} dr^2 + r^2 d \phi^2
\end{equation}
Solving the Einstein's field equation for lapse function $f(r)$ we obtain
\begin{equation}
f(r) = -M + \frac{r^2}{l^2}
\end{equation}
or equivalently, expressing the mass $M$ in terms of the horizon $r_H=l \sqrt{M}$,
\begin{equation}
f(r) = \frac{(r-r_H) (r+r_H)}{l^2}
\end{equation}
Note that the presence of the negative cosmological constant $\Lambda = -1/l^2$ is crucial for the existence of the black-hole.
We use natural units such that $c = G = \hbar = k_B = 1$ and metric signature $(-, +, +)$.

\section{Scalar perturbations: the wave equation}

Next we consider in the above gravitational background a probe scalar field with a nonzero coupling $\xi$
to the scalar curvature described by the action
\begin{equation}
S = \frac{1}{2} \int d^3x \sqrt{-g} \Bigl[  \partial^{\mu}\Phi \partial_{\mu}\Phi + \xi R_3 \Phi^2 \Bigl]
\end{equation}
In the given BTZ spacetime the wave equation of the scalar field reads \cite{coupling,kanti2,Pappas:2016ovo}
\begin{equation}
\frac{1}{\sqrt{-g}} \partial_\mu (\sqrt{-g} g^{\mu \nu} \partial_\nu) \Phi = \xi R_3 \Phi
\end{equation}
where the nonminimal coupling is taken to be positive, and $R_3=-6/l^2$ is the constant
Ricci scalar of the BTZ background. Using the ansatz
\begin{equation}\label{separable}
\Phi(t,r,\phi) = e^{-i \omega t} R(r) e^{i m \phi}
\end{equation}
we obtain an ordinary differential equation for the radial part
\begin{equation}
R'' + \left( \frac{1}{r} + \frac{f'}{f} \right) R' + \left( \frac{\omega^2}{f^2} - \frac{m^2}{r^2 f} - \frac{\xi R_{3}}{f} \right) R = 0
\end{equation}
or introducing a new mass parameter $\mu^2=6 \xi/l^2$
\begin{equation}
R'' + \left( \frac{1}{r} + \frac{f'}{f} \right) R' + \left( \frac{\omega^2}{f^2} - \frac{m^2}{r^2 f} + \frac{\mu^2}{f} \right) R = 0
\end{equation}
Note that the nonzero coupling to the scalar curvature can be interpreted as a mass term when the cosmological constant is positive. In our work however, since here the cosmological constant is negative, the mass term enters with the wrong sign.
To see the effective potential barrier that the scalar field feels we define new variables as follows
\begin{eqnarray}
R & = & \frac{\psi}{\sqrt{r}} \\
x & = & \int \frac{dr}{f(r)}
\end{eqnarray}
where we are using the so-called tortoise coordinate $x$ given by
\begin{equation}
x = \frac{l^2}{2 r_H} \ln \left( \frac{r-r_H}{r+r_H} \right)
\end{equation}
and recast the equation for the radial part into a Schr{\"o}dinger-like equation of the form
\begin{equation}
\frac{d^2 \psi}{dx^2} + (\omega^2 - V(x)) \psi = 0
\end{equation}
Therefore we obtain for the effective potential barrier the expression
\begin{equation}
V(r) = f(r) \: \left(-\frac{6 \xi}{l^2} + \frac{m^2}{r^2}+\frac{f'(r)}{2 r}-\frac{f(r)}{4 r^2} \right)
\end{equation}
Note that in the four-dimensional case the last term is absent \cite{coupling}.
The effective potential as a function of the radial distance can be seen in Fig. 1
for three different values of the coupling $\xi$.

Since the effective potential barrier vanishes at the horizon, close to the horizon $\omega^2 \gg V(x)$, and the solution for the
Schr{\"o}dinger-like equation is given by
\begin{equation}
\psi(x) = A_- e^{-i \omega x} + A_+ e^{i \omega x}
\end{equation}
Requiring purely ingoing solution \cite{kanti1, Fernando:2004ay, chinos} we set $A_-=0$ in the following.
Then the radial part close to the horizon becomes $R(r \rightarrow r_H) \sim (r-r_H)^{i \omega l^2/(2 r_H)}$.

\begin{figure}[ht!]
\centering
\includegraphics[width=\linewidth]{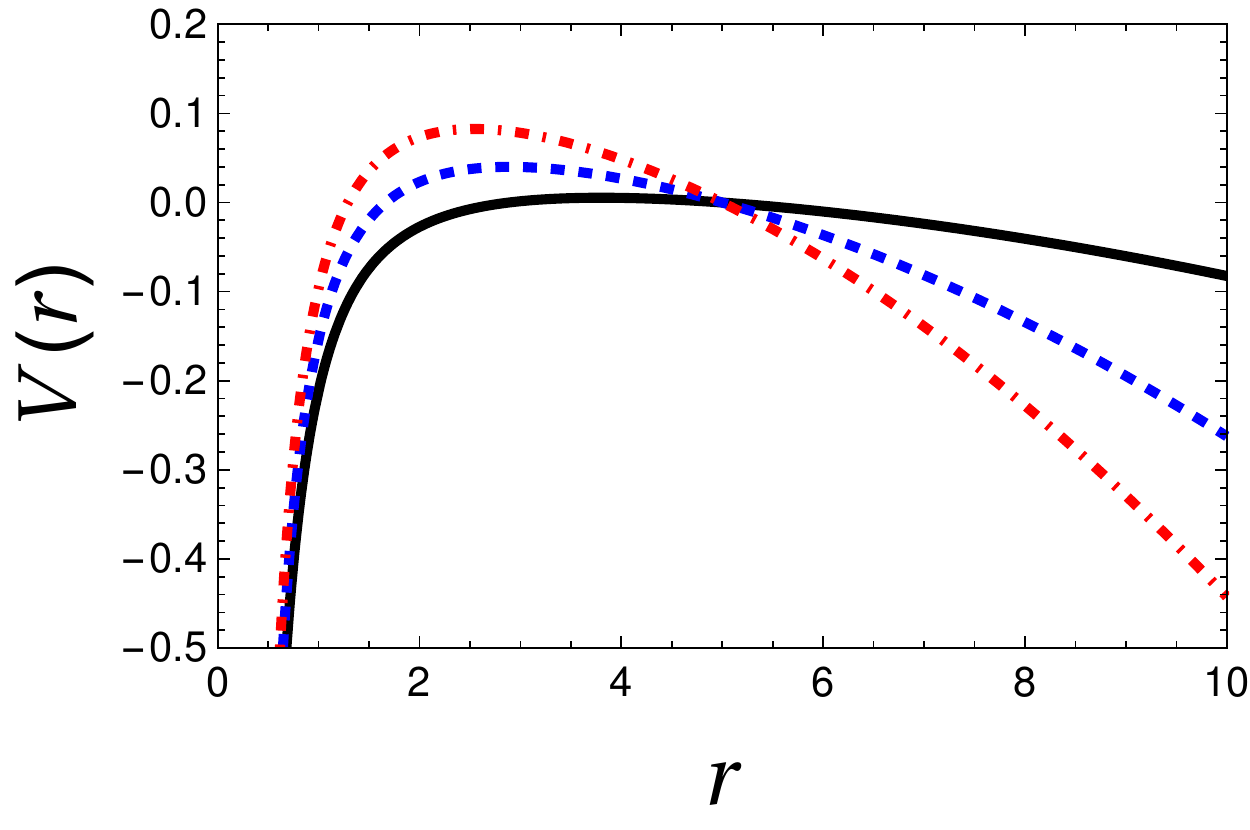}
\caption{Effective potential for $m=0, l=5, M=1$ and for $\xi=0.25$ (solid black line), $\xi=0.5$ (dashed blue line) and $\xi=0.75$ (dot-dashed red line).}
\end{figure}

\section{Solution of the radial differential equation}

\subsection{Solution in the far-field region}

To find the far field solution $r \gg r_H$ we notice that $f(r) \sim r^2/l^2$ and the mass term dominates over the frequency and angular momentum, and thus the differential equation for the radial part takes the form
\begin{equation}
r^2 R''+3 r R'+l^2 \mu^2 R = 0
\end{equation}
which is Euler's equation and it admits power-law solutions of the form $R(r) \sim r^\rho$. The power $\rho$ satisfies the algebraic equation
\begin{equation}
\rho^2 + 2 \rho + l^2 \mu^2 = 0
\end{equation}
The roots of the above equation are given by
\begin{equation}
\rho_{\pm} = -1 \pm \sqrt{1-6 \xi}
\end{equation}
which are real when $1-6 \xi \geq 0$ and complex when $6 \xi > 1$.
Therefore the far field solution is given by
\begin{equation}
R_{FF}=D_1 \left( \frac{r}{r_H} \right)^{\rho_-} + D_2 \left( \frac{r}{r_H} \right)^{\rho_+}
\end{equation}
where $D_1,D_2$ are two arbitrary coefficients. In terms of the tortoise coordinate $x$ the solution takes the form
of plane waves when the roots $\rho_{\pm}$ are complex
\begin{equation}
R_{FF} \sim x^{\pm i \sqrt{6 \xi-1}}
\end{equation}
Therefore in the following we shall consider the case where $6 \xi > 1$, and thus the roots are given by
\begin{equation}
\rho_{\pm} = -1 \pm i \sqrt{6 \xi-1}
\end{equation}
while the reflection coefficient is defined to be $\mathcal{R} = |D_1/D_2|^2$.

\subsection{Exact solution in terms of hypergeometric functions}

To find the solution of the full radial equation we introduce the dimensionless parameter $z=1-r_H^2/r^2$ which takes values
between 0 and 1, and the new differential equation becomes
\begin{equation}
z (1-z) R_{zz} + (1-z) R_z \ +
\left( \frac{A}{z} + \frac{B}{-1+z} - C \right) R = 0
\end{equation}
where the three constant are given by
\begin{eqnarray}
A & = & \frac{l^4 \omega^2}{4 r_H^2} \\
B & = & -\frac{l^2 \mu^2}{4} \\
C & = & \frac{l^2 m^2}{4 r_H^2}
\end{eqnarray}
The last differential equation can be recast in the form of the Gauss' hypergeometric equation by
removing the poles in the last term through the ansatz
\begin{equation}
R = z^\alpha (1-z)^\beta F
\end{equation}
where now $F$ satisfies the following differential equation
\begin{equation}
z (1-z) F_{zz} + [1+2 \alpha - (1+2 \alpha+2 \beta) z] F_z +
\left( \frac{\bar{A}}{z} + \frac{\bar{B}}{-1+z} - \bar{C} \right) F = 0
\end{equation}
and the new constants are given by
\begin{eqnarray}
\bar{A} & = & A + \alpha^2 \\
\bar{B} & = & B + \beta - \beta^2 \\
\bar{C} & = & C+(\alpha+\beta)^2
\end{eqnarray}
Demanding that $\bar{A}=0=\bar{B}$, we determine the parameters $\alpha$ and $\beta$ as follows
\begin{eqnarray}
\alpha & = & i \frac{l^2 \omega}{2 r_H} \\
\beta & = & \frac{1+i \sqrt{6 \xi-1}}{2}
\end{eqnarray}
and finally we obtain the hypergeometric equation
\begin{equation}
z (1-z) F_{zz} + [c-(1+a+b) z] F_z - ab F = 0
\end{equation}
with parameters $a,b,c$ given by
\begin{eqnarray}
c & = & 1+2 \alpha \\
a & = &  \alpha + \beta + i \sqrt{C} \\
b & = & \alpha + \beta - i \sqrt{C}
\end{eqnarray}
Note that the parameters $a,b,c$ satisfy the condition $c-a-b=1-2 \beta$.
Therefore the general solution for the radial part is given by \cite{chinos}
\begin{equation}
R(z) = z^\alpha (1-z)^\beta \Bigl[ C_1 F(a,b;c;z) + C_2 z^{1-c} F(a-c+1,b-c+1;2-c;z) \Bigl]
\end{equation}
where $C_1,C_2$ are two arbitrary coefficients,
and the hypergeometric function can be expanded in a Taylor series as follows \cite{handbook}
\begin{equation}
F(a,b;c;z) = 1 + \frac{a b}{c} \: z+ \cdots
\end{equation}
Setting $C_2=0$ and for the choice for $\alpha=i (l^2 \omega)/(2 r_H)$ we recover the purely ingoing solution close to the horizon,
$R \sim (r-r_H)^\alpha$.
Therefore in the following we consider the first solution only, namely
\begin{equation}
R(z) = D z^\alpha (1-z)^\beta F(a,b;c;z)
\end{equation}
where now we have replaced $C_1$ by $D$.

\subsection{Matching of the solutions}

In order to match with the far field solution obtained earlier (where now $z \rightarrow 1$) we use the transformation \cite{handbook}
\begin{equation}
\begin{split}
F(a,b;c;z) = \ &\frac{\Gamma(c) \Gamma(c-a-b)}{\Gamma(c-a) \Gamma(c-b)} \times \:
\\
&F(a,b;a+b-c+1;1-z) \ +
 \\
(1-z)^{c-a-b} &\frac{\Gamma(c) \Gamma(a+b-c)}{\Gamma(a) \Gamma(b)} \times \:
\\
&F(c-a,c-b;c-a-b+1;1-z)
\end{split}
\end{equation}
and therefore the radial part as $z \rightarrow 1$ reads
\begin{equation}
\begin{split}
R =  \frac{D (1-z)^\beta \Gamma(1+2 \alpha) \Gamma(1-2 \beta)}{\Gamma(1+\alpha-\beta-i \sqrt{C}) \Gamma(1+\alpha-\beta+i \sqrt{C})}
\\
+ \frac{ D (1-z)^{1-\beta} \Gamma(1+2 \alpha) \Gamma(-1+2 \beta)}{\Gamma(\alpha+\beta-i \sqrt{C}) \Gamma(\alpha+\beta+i \sqrt{C})}
\end{split}
\end{equation}
Note that $-2 \beta = \rho_{-}$ and $2 \beta-2=\rho_{+}$, and since $z=1-(r_H/r)^2$ the radial part $R(r)$ for $r \gg r_H$ can be written down as follows
\begin{equation}
\begin{split}
R =  \frac{D \Gamma(1+2 \alpha) \Gamma(1-2 \beta) \left(\frac{r}{r_H}\right)^{\rho_{-}} }{\Gamma(1+\alpha-\beta-i \sqrt{C}) \Gamma(1+\alpha-\beta+i \sqrt{C})}
 \\
+ \frac{ D \Gamma(1+2 \alpha) \Gamma(-1+2 \beta) \left(\frac{r}{r_H}\right)^{\rho_{+}} }{\Gamma(\alpha+\beta-i \sqrt{C}) \Gamma(\alpha+\beta+i \sqrt{C}) }
\end{split}
\end{equation}
Finally upon comparison we express $D_1, D_2$ in terms of $D$ as follows
\begin{eqnarray}
D_1 & = & D \: \frac{\Gamma(1+2 \alpha) \Gamma(1-2 \beta)}{\Gamma(1+\alpha-\beta-i \sqrt{C}) \Gamma(1+\alpha-\beta+i \sqrt{C})} \\
D_2 & = & D \: \frac{\Gamma(1+2 \alpha) \Gamma(-1+2 \beta)}{\Gamma(\alpha+\beta-i \sqrt{C}) \Gamma(\alpha+\beta+i \sqrt{C})}
\end{eqnarray}

\section{Numerical results}

We can now present the analytical expressions and discuss the numerical results.

\subsection{The reflection coefficient}

\begin{figure}[ht!]
\centering
\includegraphics[width=\linewidth]{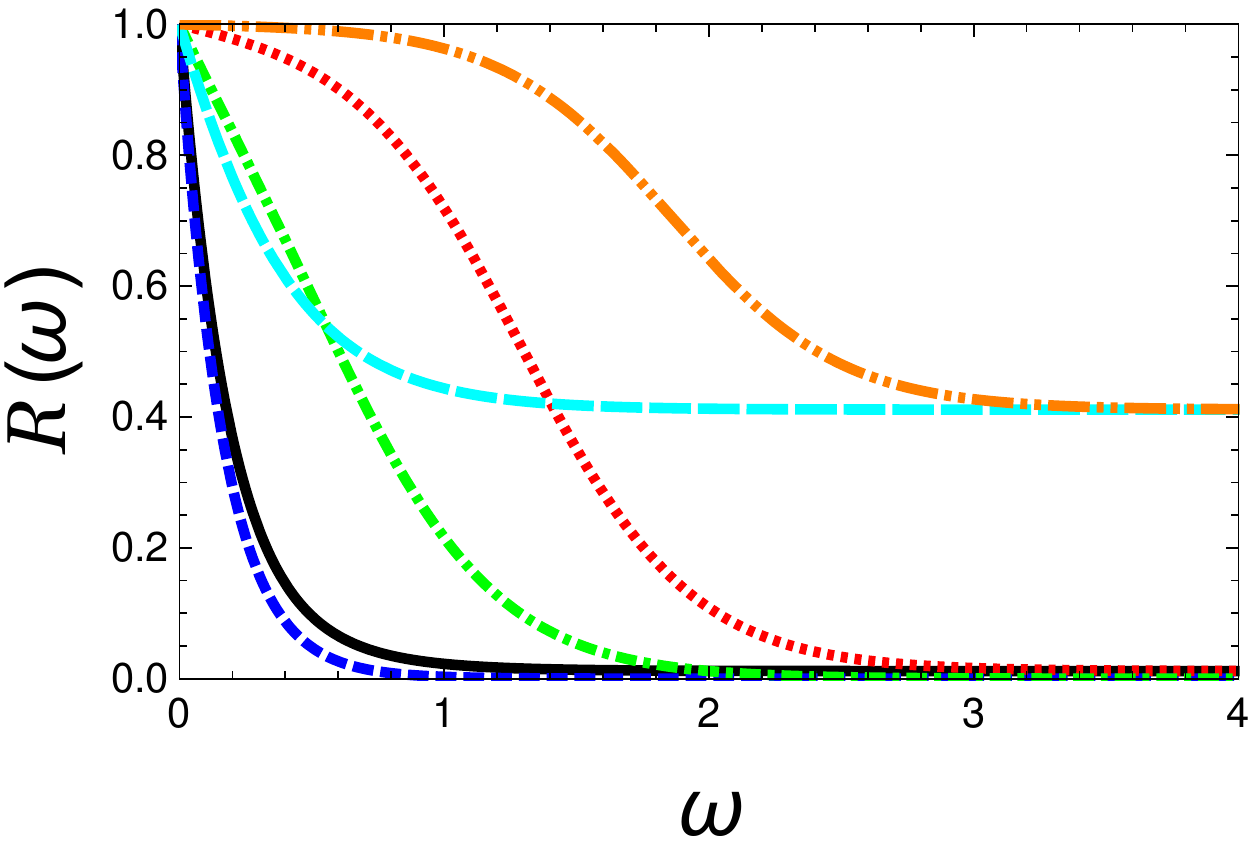}
\caption{\label{Reflection}
Reflection coefficient taking $l=M=1$ for $m=0, \ \xi=0.17$ (long dashed cyan line),
$m=0,\ \xi=0.25$ (solid black line), $m=0,\ \xi=0.5$ (short dashed blue line),
$m=2, \ \xi=0.17$ (double dotted dashed orange line),
$m=2, \ \xi=0.25$ (dotted red line) and $m=2,\ \xi=0.5$ (dotted dashed green line).}
\end{figure}

First, the reflection coefficient defined to be $\mathcal{R} = |D_1/D_2|^2$ can be calculated using the following identities
for the $\Gamma$ function \cite{Fernando:2004ay}
\begin{eqnarray}
|\Gamma(i y)|^2 & = & \frac{\pi}{y \sinh(\pi y)}  \\
\bigg|\Gamma\left(\frac{1}{2}+i y\right)\bigg|^2 & = & \frac{\pi}{\cosh(\pi y)}
\end{eqnarray}
and we obtain the expression
\begin{equation}\label{R1}
\begin{split}
\mathcal{R} = \frac{\cosh\left[\frac{\pi}{2} \bigg(\frac{l^2 \omega}{r_H}-\sqrt{6 \xi-1}-\frac{l m}{r_H}\bigg)\right]
}{\cosh\left[\frac{\pi}{2} \bigg(\frac{l^2 \omega}{r_H}+\sqrt{6 \xi-1}-\frac{l m}{r_H}\bigg)\right]
} \times
\\
\frac{\cosh\left[\frac{\pi}{2} \bigg(\frac{l^2 \omega}{r_H}-\sqrt{6 \xi-1}+\frac{l m}{r_H}\bigg)\right]}{\cosh\left[\frac{\pi}{2} \bigg(\frac{l^2 \omega}{r_H}+\sqrt{6 \xi-1}+\frac{l m}{r_H}\bigg)\right]}.
\end{split}
\end{equation}
The reflection coefficient as a function of the frequency can be seen in Fig. \ref{Reflection}. We have considered six cases, namely $\xi=0.17$, $\xi=0.25$ and $\xi=0.5$ for vanishing angular momentum $m=0$, and $\xi=0.17$, $\xi=0.25$ and $\xi=0.5$ for non-vanishing angular momentum $m=2$. We can observe the following features: i) For $m \neq 0$ the reflection coefficient decays slower than for $m = 0$, ii) For non-vanishing angular momentum the curvature of the curves changes sign at $\omega_*=m/l$, and iii) $\mathcal{R}$ starts at 1, it monotonically decreases and finally tends to zero in the high energy regime only for sufficiently high values of the coupling $\xi \geq 0.25$. On the contrary, when $\xi$ takes values relatively close to its conformal value, $\mathcal{R}$ monotonically decreases to a finite constant different than zero. Therefore in the rest of the discussion we shall only consider high values of $\xi$. To be more precise, we have checked that in the high energy regime
\begin{equation}
\mathcal{R}(\omega \rightarrow \infty) \sim e^{-2 \pi \sqrt{6 \xi-1}}
\end{equation}
irrespectively of the rest of the parameters, and therefore the reflection coefficient really tends to zero only in the $\xi \rightarrow \infty$ limit.

\subsection{The absorption cross section}

\begin{figure}[ht!]
\centering
\includegraphics[width=\linewidth]{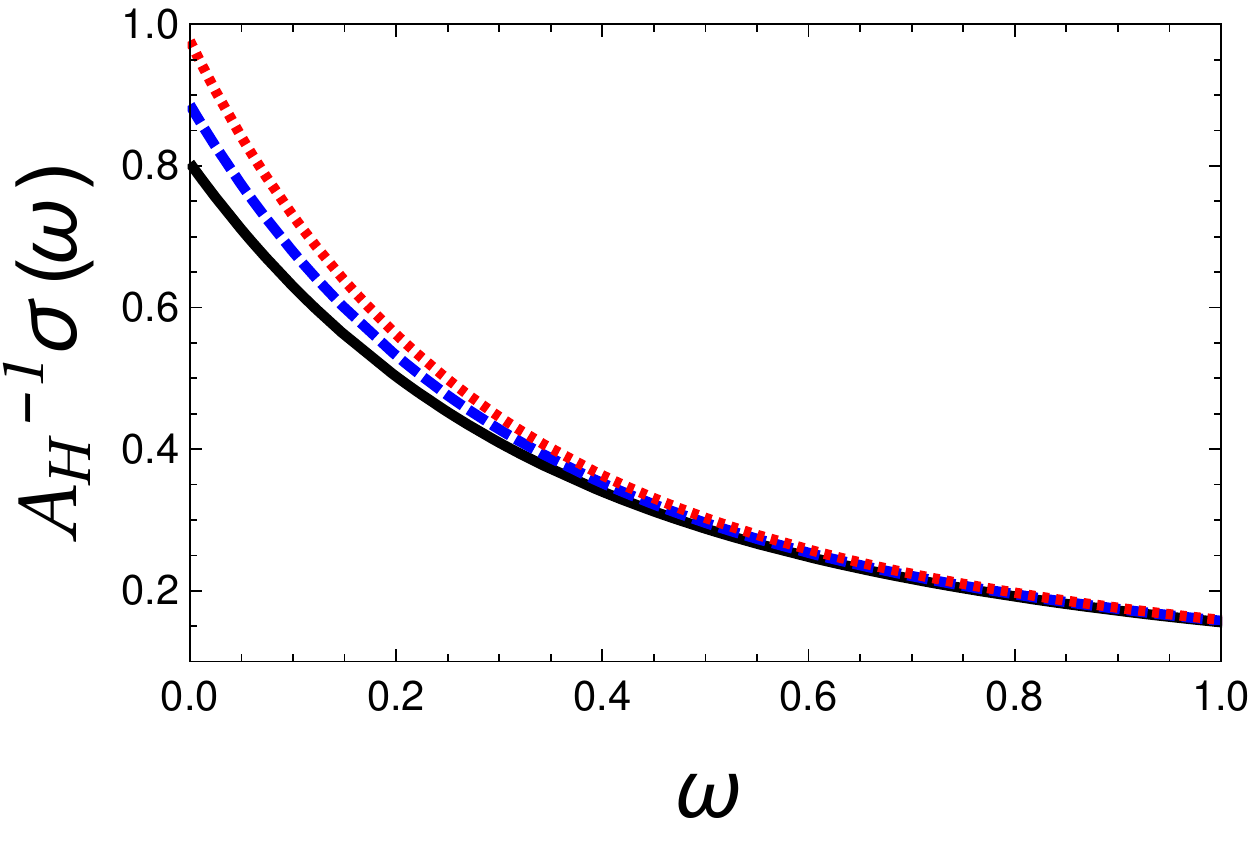}
\caption{\label{Ratio2}
Absorption cross-section (in units of the horizon area) for $m=0$ and $l=M=1$ for $\xi=0.25$ (black solid line), $\xi=0.3$ (dashed blue line) and $\xi=1$ (dotted red line).}
\end{figure}

\begin{figure}[ht!]
\centering
\includegraphics[width=\linewidth]{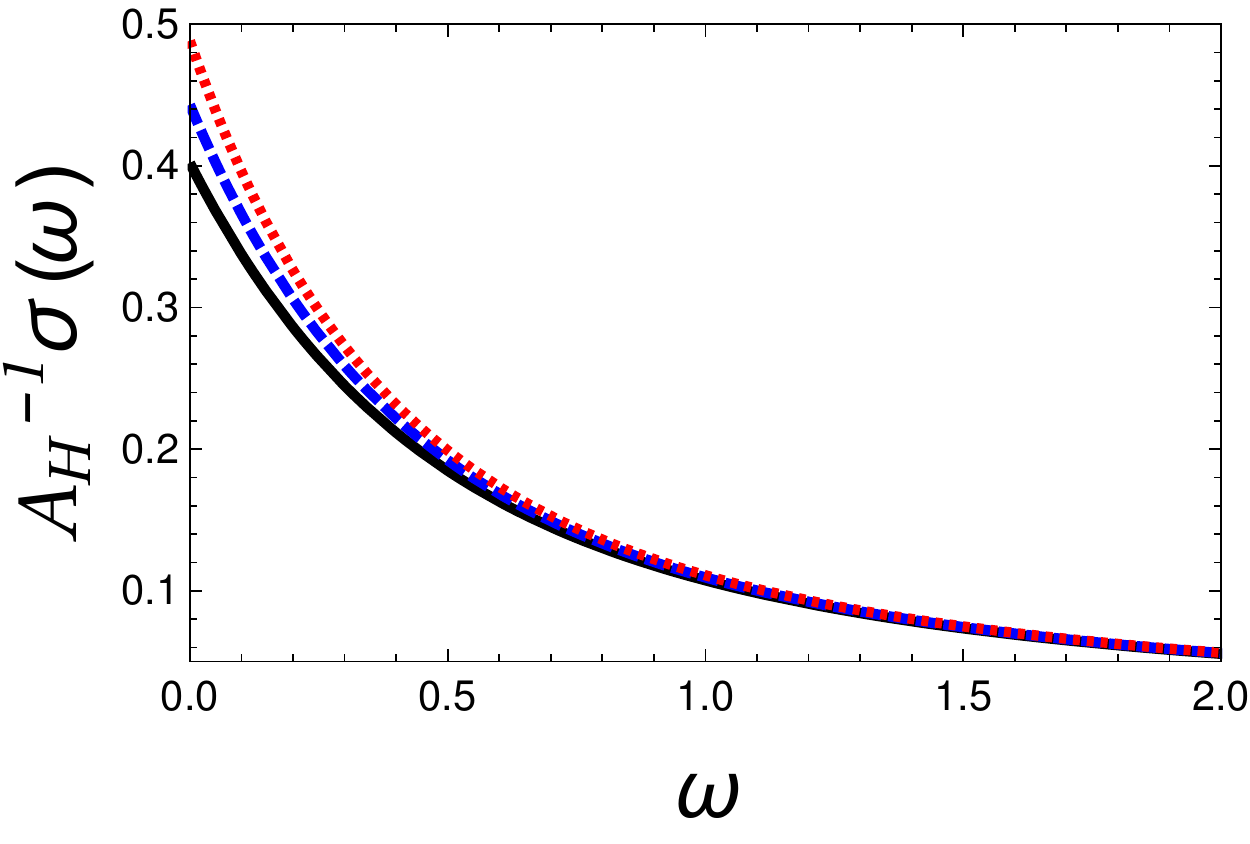}
\caption{\label{Ratio1}
Absorption cross-section (in units of the horizon area) for $m=0$ and $l=1, M=2$ for $\xi=0.25$ (black solid line), $\xi=0.3$ (dashed blue line) and $\xi=1$ (dotted red line).}
\end{figure}

Then the absorption cross section is given by the simple formula \cite{Fernando:2004ay,universality}
\begin{equation}
\sigma_{abs} = \frac{1-\mathcal{R}}{\omega}
\end{equation}
and in the low energy regime it tends to a finite constant. This can be seen if we Taylor expand $\mathcal{R}(\omega)$ around $\omega = 0$ obtaining
\begin{equation}
\mathcal{R}(\omega) = 1-\sigma_0 \omega + \mathcal{O}(\omega^2)
\end{equation}
and thus $\sigma_{abs}(\omega \rightarrow 0) \rightarrow \sigma_0$, where the constant $\sigma_0$ is given by
\begin{equation}
\sigma_0 = \frac{\pi l}{\sqrt{M}} \Bigg[ \tanh\bigg[\frac{\pi m}{2 \sqrt{M}}+\frac{\pi \sqrt{6 \xi-1}}{2}\bigg] - \tanh\bigg[\frac{\pi m}{2 \sqrt{M}}-\frac{\pi \sqrt{6 \xi-1}}{2}\bigg] \Bigg]
\end{equation}
In \cite{universality} it was shown that for a generic spherically symmetric black hole the absorption cross section of a minimally coupled
massless scalar field in the low energy regime tends to a constant that coincides with the area of the horizon.
It is not obvious that this is still true for a nonminimally scalar field. In fact in \cite{coupling} it was shown that for a four-dimensional
nonminimally coupled scalar field the absorption cross-section in the low energy regime tends to zero like $\omega^2$. In this work we find
that as a function of $\omega$ for $m=0$ the absorption cross-section starts from a constant and eventually goes to zero, but this constant
does not necessarily coincide with the area of the horizon $\mathcal{A}_H=2 \pi l \sqrt{M}$. We define the dimensionless
parameter $\sigma_{abs}/\mathcal{A}_H$ and we plot it
as a function of the frequency in Fig. \ref{Ratio2} and \ref{Ratio1} for the case where $l=r_H$ ($M=1$) and for the case in which $r_H > l$ ($M > 1$).
The constant increases with the coupling and finally acquires a limiting value when the coupling becomes sufficiently large.
When $M=l$ (Fig. \ref{Ratio2}) this limiting value is precisely one, while when $M > l$ (Fig. \ref{Ratio1}) the limiting value remains always below unity.
This can be seen using the Taylor expansion of $\mathcal{R}(\omega)$ obtained earlier. When $m=0, \omega \rightarrow 0$ the absorption cross section per area
takes the simple form
\begin{equation}
\frac{\sigma_{abs}}{\mathcal{A}_H} = \frac{1}{M} \tanh\left(\frac{1}{2}\pi \: \sqrt{6 \xi-1}\right)
\end{equation}
and it does not depend on the cosmological constant. Furthermore, when $\xi \gg 1$
it becomes $1/M$, and therefore it is 1 only when $M=1$ ($l=r_H$), while it remains
smaller than 1 when $M > 1$ ($r_H > l$).

\subsection{The decay rate}

\begin{figure}[ht!]
\centering
\includegraphics[width=\linewidth]{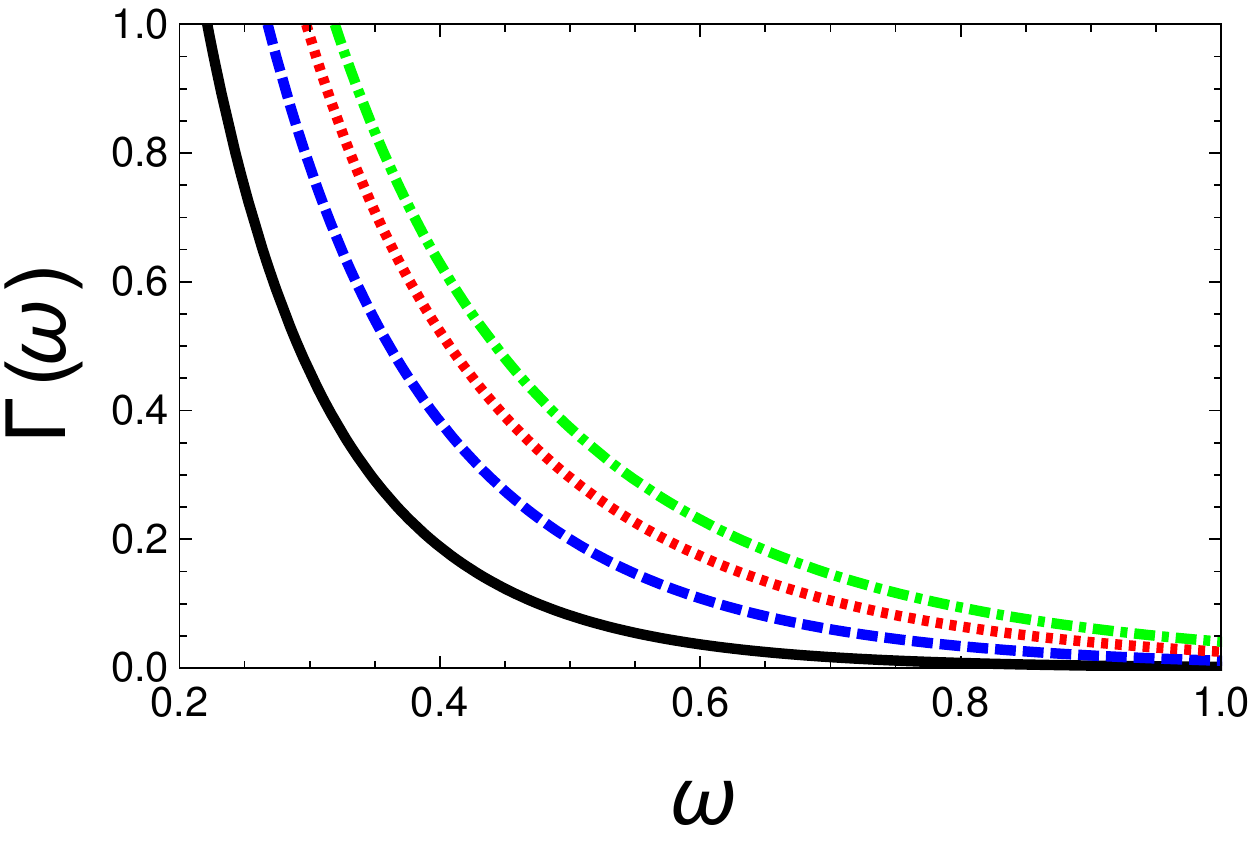}
\caption{\label{RateI}
Decay rate for $l=1$, $\xi = 0.25$, $m=0$ and for $M = 1$ (solid black line), $M = 2$ (dashed blue line), $M = 3$ (dotted red line) and $M=4$ (dot-dashed
green line).}
\end{figure}

\begin{figure}[ht!]
\centering
\includegraphics[width=\linewidth]{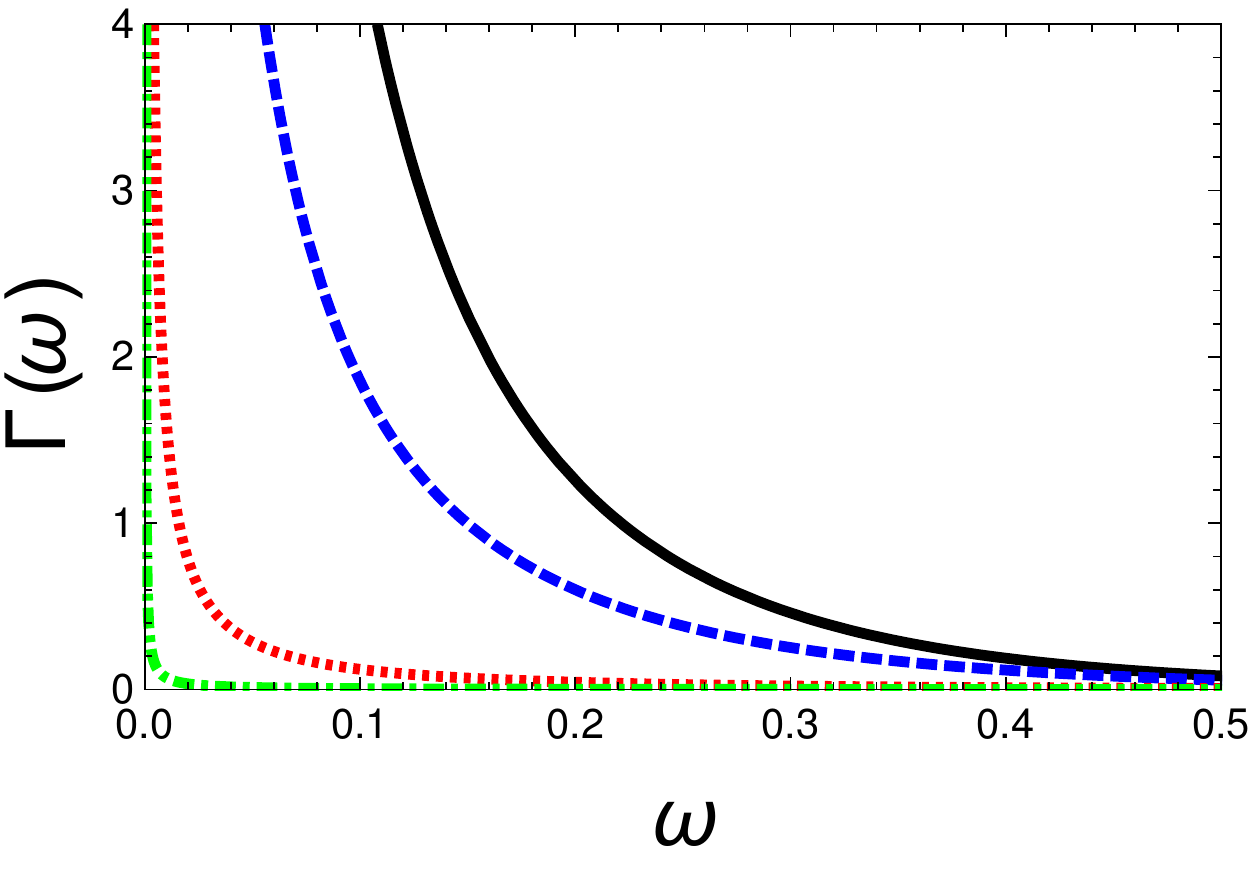}
\caption{\label{RateII}
Decay rate for $l=1$, $\xi = 0.25$, $M = 1$ and for $m = 0$ (solid black line), $m = 1$ (dashed blue line) and $m = 2$ (dotted red line) and $m=3$ (dot-dashed green line).}
\end{figure}

Finally, since the flux spectrum  emitted by the black hole is given by \cite{coupling}
\begin{equation}\label{flux}
\frac{dN(\omega)}{dt} = \sum_{\ell} \frac{\sigma_{\ell}(\omega)}{e^{\omega/T_H} - 1} \frac{d^3 k}{(2 \pi)^3}
\end{equation}
we define the decay rate for the black hole to by \cite{Fernando:2004ay}:
\begin{align}
\Gamma_{decay} = \frac{\sigma_{abs}}{e^{\omega/T_H} - 1}
\end{align}
where the Hawking temperature of the BTZ black hole is given by $T_H=r_H/2 \pi l^2$ or $T_H=\sqrt{M}/(2 \pi l)$ \cite{hawking2}. Hence
\begin{align}
\Gamma_{decay} &=
\frac{ \frac{1}{\omega}\sinh \left(\pi  \sqrt{6 \xi -1}\right) e^{-\frac{\pi  l^2 \omega }{r_H}}}{ \cosh \left[\pi  \left(\frac{l^2 \omega }{r_H}+\sqrt{6 \xi -1}\right)\right] +\cosh \left(\frac{\pi  l m}{r_H}\right)}
\end{align}
The decay rate as a function of frequency can be seen in the Fig. \ref{RateI} and Fig. \ref{RateII}. In both figures it tends to
$\infty$ when $\omega$ tends to zero, and monotonically decreases to zero. The more important effects are parameterized by $m$ and $M$,
however the nonminimal coupling constant $\xi$ also affects slightly the behavior. Our figures are consistent with those of \cite{Fernando:2004ay}
in the high energy regime: the curves asymptotically go to zero.
We observe in Fig. \ref{RateI} that when we increase the mass of the black hole, the decay rate decreases more slowly than for smaller masses. In a similar way, the effect
of the angular momentum on the behavior of the decay rate is shown in Fig. \ref{RateII}. Thus, cases
for $m \neq 0$ imply a decay rate that goes to zero faster compared to the case of $m=0$.

\section{Conclusions}

To summarize, in this article we have analyzed the propagation and relativistic scattering of a probe scalar field with a nonminimal coupling to gravity in BTZ black hole background. We have found analytical expressions for the reflection coefficient, the absorption cross-section, and the decay rate as functions of the frequency in the strong coupling case $\xi > \xi_{c}=1/6$, and we have shown in figures how these quantities depend on the parameters of the theory. According to the Universality theorem, in the zero angular-momentum case and in the low energy regime, the greybody factor for a minimally
coupled massless scalar field tends to a finite constant that coincides with the area of the black-hole horizon. However, it is expected that this will not be true in general neither for a nonminimally coupled scalar field nor for a massive scalar field. Our results show that in the zero angular-momentum case the greybody factor in the low energy regime tends to a constant, contrary to the case of a nonminimally coupled scalar field in a four-dimensional
spacetime with a positive cosmological constant, where the greybody factor tends to zero in the zero-frequency limit
like $\omega^2$. This finite constant generically does not coincide with the area of the horizon. There is, however, a special case where this holds, and this happens only when $M=1$. A nonzero coupling to the scalar curvature looks like a mass term, however
for a negative cosmological constant the mass terms enters with the wrong sign. In spite of that, our results resemble the results of a previous work \cite{massive} in which the authors found that in the case of a massive scalar field, Universality is respected only under certain restrictions of the parameters involved.


\section*{Acknowlegements}

We wish to thank the anonymous reviewer for useful comments and suggestions.
The author G.P. acknowledges the support from "Funda{\c c}{\~a}o para a Ci{\^e}ncia e Tecnologia". The author A.R. was supported
by the CONICYT-PCHA/\- Doctorado Nacional/2015-21151658. 



\end{document}